\documentclass{aastex}
\usepackage{emulateapj5}
\input psfig

\newcommand\lsun{{\rm\,L_\odot}}

\submitted{Astrophysical Journal Accepted, 22 October 2001}
\shortauthors{Nelson et al.}
\shorttitle{Constraints on Size Evolution of Brightest Cluster Galaxies}

\def\littleprime{\ifmmode{\scriptscriptstyle \prime }   
\else{\hbox{$\scriptscriptstyle \prime$ }}\fi}

\def\arcsec{\raise .9ex \hbox{\littleprime\hskip-3pt\littleprime}}
\def\arcmin{\raise .9ex \hbox{\littleprime}}
\def\arcsecpoint{\hbox to 1pt{}\rlap{\arcsec}.\hbox to 2pt{}}
\def\arcminpoint{\hbox to 1pt{}\rlap{\arcmin}.\hbox to 2pt{}}

\def\spose#1{\hbox to 0pt{#1\hss}}
\def\lta{\mathrel{\spose{\lower 3pt\hbox{$\mathchar"218$}}
     \raise 2.0pt\hbox{$\mathchar"13C$}}}
\def\gta{\mathrel{\spose{\lower 3pt\hbox{$\mathchar"218$}}
     \raise 2.0pt\hbox{$\mathchar"13E$}}}

\begin{document}
\input{psfig.sty}

\title{Constraints On the Size Evolution of Brightest Cluster Galaxies\altaffilmark{1}}
\author{Amy E. Nelson}
\affil{Board of Astronomy and Astrophysics, University of California, Santa
Cruz, CA, 95064, E-Mail: anelson@ucolick.org}

\author{Luc Simard\altaffilmark{2}, Dennis Zaritsky} \affil{Steward Observatory, 933 N. Cherry
Ave., University of Arizona, Tucson, AZ, 85721, E-Mail:
lsimard, dzaritsky@as.arizona.edu}  

\author{Julianne J. Dalcanton}
\affil{Box 351580, University of Washington, Seattle, WA, 98195, E-Mail:
jd@toast.astro.washington.edu}

\author{and}

\author{Anthony H. Gonzalez}
\affil{Harvard-Smithsonian Center for Astrophysics, 60 Garden Street,
Cambridge, MA, 02138, E-Mail: anthony@head-cfa.harvard.edu}

\altaffiltext{1}{Based on observations made with the NASA/ESA {\it Hubble Space Telescope} which is operated by AURA, Inc., under contract with NASA.}

\altaffiltext{2}{Guest User, Canadian Astronomy Data Center, which is 
operated by the National Research Council of Canada, Herzberg 
Institute of Astrophysics, Dominion Astrophysical Observatory.}

\begin{abstract}
We measure the luminosity profiles of 16 brightest cluster galaxies
(BCGs) at $0.4 < z < 0.8$ using high resolution F160W NICMOS and F814W
WFPC2 HST imaging. The heterogeneous sample is drawn from a variety of
surveys: seven from clusters in the Einstein Medium Sensitivity Survey
\citep[EMSS;][]{gioia94}, five from the Las Campanas Distant Cluster
Survey and its northern hemisphere precursor \citep[LCDCS;][]{dal97,gonzalez01,nelson01a}, and the remaining four from
traditional optical surveys \citep{spinrad80,koo81,gunn86,couch91}.  We find that the surface brightness
profiles of all but three of these BCGs are well described by a standard
de Vaucouleurs ($r^{1/4}$) profile out to at least $\sim2r_{e}$ and
that the biweight-estimated NICMOS effective radius of our high
redshift BCGs ($r_{e} = 8.3\pm1.4$ kpc for $H_{0} = 80$ km s$^{-1}$
Mpc$^{-1}$, $\Omega_{m} = 0.2, \Omega_\Lambda = 0.0$) is $\sim 2$ times
smaller than that measured for a local BCG sample
\citep{graham96}. If high redshift BCGs are in dynamical equilibrium
and satisfy the same scaling relations as low redshift ones, this
change in size would correspond to a mass growth of a factor of 2
since $z \sim 0.5$.  However, the biweight-estimated WFPC2 effective
radius of our sample is 18 $\pm $ 5.1 kpc, which is fully consistent
with the local sample.  While we can rule out mass accretion rates higher
than a factor of 2 in our sample, the discrepancy between our NICMOS
and WFPC2 results, which after various tests we describe appears to be
physical, does not yet allow us to place strong constraints on
accretion rates below that level.  
\end{abstract}


\section{Introduction}\label{intro}

Brightest cluster galaxies (BCGs) have been studied, both
photometrically and spectroscopically, in great detail at low redshift
because of their unique characteristics and location
\citep[e.g.,][]{hoessel80,schombert86,postman95}. BCGs
are the most massive and luminous galaxies in the universe, with
central velocity dispersions $\sigma \sim 300-400$ km s$^{-1}$
\citep{dressler79,carter85,fisher95}, and typical
luminosities $\sim 10$ L$_{*}$, where L$_{*} = 1.0 \times
10^{10}h^{2}\lsun$, \citep{sandage73,hoessel80,schombert86}. Rather than 
simply being ``the brightest of the bright," BCGs are drawn from a different luminosity function
\citep{dressler78} and follow tighter scaling relations, for example
the $r_{e}-\mu_{e}$ relation \citep{hoessel87}, than other cluster
ellipticals. Their remarkably homogeneous luminosity, $\sigma_{V}
\simeq 0.35$ mag, ($\sigma_{V} \simeq 0.30$ mag when applying
corrections based upon environment; \citealp[e.g., ][]{sandage73,hoessel80}), 
suggests they may experience a unique evolutionary
history. Because BCGs are typically located near the center of the
cluster potential, they may preferentially accrete material from
tidally stripped cluster galaxies \citep{richstone76} or merge with
other cluster members \citep{ostriker77}.  Observed correlations
between BCG structural parameters and luminosity (such as the
luminosity-shape relation of Hoessel et al. 1980) may be signs of such
accretion events \citep{ostriker75,schneider83}.

Studies comparing the luminosity of high redshift BCGs to their low
redshift counterparts find indirect evidence of mass accretion in BCGs
since $z \sim 1$ (\citealp[][hereafter ABK98]{aragon98}; \citealp{burke00,nelson01a}).  
The amount of inferred evolution, however, is tied to environment. BCGs in clusters with low x-ray luminosity ($L_{x} <
2.3 \times 10^{44}$ ergs s$^{-1}$) fade less with decreasing redshift
than expected from the aging of their stellar populations \citep[ABK98;][]{burke00,nelson01a}. The proposed explanation for this
observation is that these galaxies are accreting roughly enough mass
in stars to counter the expected passive evolution -- about a factor
of 2 to 4 increase in mass since $z \sim 1$ (ABK98). On the other
hand, the luminosities of high $L_{x}$ BCGs do match the predictions
of passive evolution models \citep{burke00,nelson01a}, which
limits the possible amount of mass these BCGs accrete to less than a
factor of 2.

Unfortunately, the inferred mass accretion rates rely on the
assumption that increases in luminosity are attributable to increases
in mass. A way to test the hypothesis of accretion in BCGs is to
measure the structural parameters of the BCGs (mass, size, velocity
dispersion, surface brightness). Locally, BCGs conform to the
fundamental plane relations between log $r_{e}$$-$ log $\sigma-
\langle\mu\rangle$ derived by \citet{dressler87} for lower
luminosity ellipticals \citep{oegerle91}. Because they lie on the
fundamental plane, the local BCGs appear to be in dynamical
equilibrium and changes in mass should produce corresponding changes
in size. If we presume that high redshift BCGs also satisfy these
scaling relations on the grounds that high redshift ellipticals are
observed to lie on the fundamental plane \citep{kelson97}, then a
comparison of the sizes of local and distant BCGs should help
constrain accretion models.  Alternatively, highly irregular BCG
luminosity profiles would argue that these galaxies are not in
dynamical equilibrium. The major obstacle for such a study is that
measuring structural parameters of high redshift BCGs requires high
angular resolution imaging.

We present the first study of the structural parameters of high
redshift BCGs using {\it Hubble Space Telescope} NICMOS F160W and
WFPC2 F814W imaging and search for direct signatures of accretion
events in these galaxies. The use of infrared data is important
because, even for our most distant clusters, these wavelengths probe
the older, more quiescent stellar populations, thereby making
comparisons to local samples more direct.  Without prior knowledge, a
homogeneous data set that minimizes one's sensitivity to potential
recent star formation is necessary for this study. If the
high-redshift BCGs are found to be quiescent, relaxed systems, then
one might, with caution, further explore the WFPC2 archive to enlarge
the sample. In \S2 we describe our heterogeneous sample of 16 BCGs at
$0.4 < z < 0.8$ and the HST data. In \S3, we describe the three
different methods we use to measure BCG effective radii including
GIM2D, a package which we utilize to fit the surface brightness
profiles of the BCGs. In \S4, we present the best fitting models to
the surface brightness profiles and investigate the mass accretion
rate of BCGs since $z \sim 0.5$.  Finally, we summarize our conclusions
in \S5.

\section{The Data}\label{data}

\subsection{HST/NICMOS}\label{data-nic2} The 16 BCGs we analyze are a
heterogeneous sample culled from several sources -- seven from
clusters in the Einstein Medium Sensitivity Survey \citep[EMSS;][]{gioia94}, 
five from the Las Campanas Distant Cluster Survey and
its northern hemisphere precursor \citep[LCDCS;][]{dal97,gonzalez01,nelson01a}, 
and the remaining four from traditional optical surveys 
\citep{spinrad80,koo81,gunn86,couch91}. The BCGs were observed by the Hubble
Space Telescope with the NICMOS camera and F160W filter from June 1997
through March 1998 in the ``snapshot'' mode as part of program GO 7327
(PI: Dalcanton).  The images were obtained using the NIC2 camera
configuration ($19\arcsecpoint2 \times 19\arcsecpoint2$ field
of view, 0.075$^{\prime\prime}$ pixel$^{-1}$) in MULTIACCUM mode with
NSAMP=16. Observations consist of three dithered exposures per object
with exposure times of 256 seconds per dither position for BCGs at $z
< 0.6$ and 352 seconds per dither position for BCGs at $z > 0.6$.  The
NIC2 image datasets are listed in Table~\ref{hst-data}.  NICMOS images
calibrated ``on-the-fly'' from the HST data archive are generally of
inferior quality because the automated reduction pipeline, CALNICA,
does not correct for non-linear bias levels and time-varying changes
in the dark current level (the so-called ``pedestal effect'').
Therefore, we calibrate the raw science images using the NICPIPE
routine available in the STSDAS Version 2.2 IRAF package. This
pipeline incorporates the standard reduction procedure of CALNICA, but
in addition uses the tasks BIASEQ and PEDSUB to correct for non-linear
bias level drift and quadrant-dependent residual bias level.

We calculate F160W magnitudes (denoted $H_{160, AB}$ hereafter) in
Oke's AB system \citep{oke83} according to the following expression:

\begin{equation} H_{160,AB} = -2.5 log(PHOTFNU \times CR) + 8.9,
\end{equation}

\noindent where $CR$ is the count rate and $PHOTFNU$ is the $HST$
photometric conversion between countrates and fluxes in units of Jy s
DN$^{-1}$ in the image headers. Galaxy magnitudes are corrected for
extinction using the dust IR emission maps of \citet{schlegel98}
but are not K-corrected.

The galaxy identified as the BCG was generally not known prior to this
program.  Because the field of view of the NICMOS images is small, we
do not simply choose the brightest galaxy on the image as the
BCG. Instead, we identify the BCGs on large optical and/or near-IR
images for all but two clusters -- MS0302.5$+$1717 and MS1333.3$+$1725
\citep{gioia94}. For the LCDCS clusters, the BCG is defined as the
galaxy with the brightest total $I$-band magnitude located within $350
h^{-1}$ kpc of the cluster center. To reduce contamination by bright
blue foreground galaxies, we exclude BCG candidates whose colors are
0.4 mag bluer than the location of the red envelope in the cluster
color-magnitude diagram (see \citealt{nelson01a} for details).
1100$+$4620 does not have $I$-band imaging and therefore the BCG is
determined using ground-based $Kp$ imaging. We identify the BCGs in
CL0016$+$16, J1888.16CL, CL0317$+$1521, and CL1322$+$3027 using the
cluster contour plots and $K$-band photometry files published by
\citet{aragon93}. Five of the seven EMSS BCGs are
selected using various optical images and identifications published
in the literature \citep{luppino92,clowe98,luppino99,clowe00}. 
Finally, we could not find BCG identifications in the
literature for the remaining two EMSS clusters (MS0302.5$+$1717 and
MS1333.3$+$1725). The NICMOS images for both of these clusters contain
a large, dominant galaxy in the field of view which we define to be
the BCG. We caution, however, that the BCG identifications for these
two clusters are less certain.

\subsection{HST/WFPC2}\label{data-wfpc2} We also make use of archival
HST/WFPC2 images taken in the F814W filter available for 9 BCG's in
our NICMOS sample. The datasets are listed with their exposure times
in Table~\ref{hst-data}.  These archival images were recalibrated
``on-the-fly" through the Canadian Astronomy Data Center (CADC)
standard pipeline. WFPC2 images covers a larger field of view
(36\arcsecpoint8 $\times$ 36\arcsecpoint8 for the Planetary Camera and
80\arcsec $\times$ 80\arcsec for each of the three Wide-Field Camera detectors) than
NIC2. At $z = 0.6$, the F814W filter samples the rest-frame wavelength
range 4578$-$5676 \AA. As a comparison, the NIC2 F160W filter at that
redshift samples the rest-frame wavelength range 8750$-$11250 \AA.

\section{Analysis}\label{analysis} One of the main problems in
measuring the sizes of high-redshift brightest cluster galaxies are
the systematic errors in the measurement of the sky. This
problem is exacerbated by the fact that some BCGs locally exhibit
steep (S\'ersic index $n > 4$) profiles and a significant amount of
flux in their broad outer wings \citep{graham96}. Sky errors are of
particular concern for the NIC2 detector because it covers a
relatively small sky area, which can lead to contamination of
flat-fielding and sky subtraction.  Because of the uncertainties
involved in the sky determination, we decided to measure BCG sizes in
three different ways: curve of growth with circular apertures (Section
\ref{analysis-cog}), Point-Spread-Function (PSF)-convolved surface
brightness models (Section \ref{analysis-gim2d}), and isophotal
ellipse fitting (Section \ref{analysis-isoellipse}).

\subsection{Curve of Growth}\label{analysis-cog} For each BCG in our
sample, we measure NIC2 and WFPC2 circular half-light radii by
computing the galaxy flux within concentric apertures and
interpolating this curve of growth to find the radius that contains
half of the total flux given by the asymptotic flux value at large
radii.  The curves of growth were measured directly from the images 
and not from surface brightness models such as the ones obtained in 
Section~\ref{analysis-gim2d}. Typical maximum radii for the NIC2 and 
WFPC2 curves of growth are 6.8$-$9.0 arcseconds (90$-$120 pixels) and 
20 arcseconds (200 pixels) respectively.  The sky background is manually 
adjusted until the integrated flux is asymptotically flat over a range of at least 30
pixels. Flux from neighboring galaxies is carefully excluded from the
curves of growth using segmentation images produced with the
SExtractor galaxy photometry package \citep{bertin96}. The actual BCG
flux in each excluded pixel, which would have been included in the
curve of growth in the absence of neighbors, is calculated using its
symmetric counterpart about the center of the BCG. The NIC2
curve-of-growth BCG half-light radii (in arcseconds and in kpc) is
presented in Table~\ref{nic2-sizes} (Columns 3 and 4). We list three
values for each BCG because a different curve of growth was obtained
independently for each of the three NIC2 dither positions. The typical
variations in half-light radii from one dither position to another are
$\sim$ 0\arcsecpoint1$-$0\arcsecpoint2. The WFPC2 curve-of-growth
half-light radii are presented in Table~\ref{wfpc2-sizes}.

\subsection{BCG Surface Brightness Profile
Fitting}\label{analysis-gim2d}

The BCG surface brightness profiles are fit using the GIM2D package,
which is designed to perform 2D disk+bulge deconvolutions of low
signal-to-noise images of distant galaxies \citep{simard98,simard01}.

The GIM2D bulge component profile is a S\'ersic profile
\citep{sersic68} of the form:

\begin{equation} \Sigma(r) = \Sigma_{e} {\rm
exp}\{-b[({{r}/{r_{e}}})^{1/n}-1]\}, \label{sersic-prof}
\end{equation}

\noindent where $\Sigma(r)$ is the surface brightness along the
semimajor axis at radius $r$, $r_{e}$ is the semi-major axis effective
radius, and $\Sigma_{e}$ is the effective surface brightness. The
parameter $b$ is set equal to $1.9992n-0.3271$ to ensure that $r_{e}$
is the projected radius enclosing half of the light. A standard de
Vaucouleurs profile is obtained by setting $n = 4$. The flux interior
to the radius $r$ is given by

\begin{equation} F(r) =2\pi
n\Sigma_{e}r_{e}^{2}{{e^{b}}\over{b^{2n}}}\gamma(2n,b),
\label{sersic-flux} \end{equation}

\noindent where $\gamma(2n,x)$ is the incomplete Gamma function with
$x = b(r/r_{e})^{1/n}$. Equations \ref{sersic-prof} and
\ref{sersic-flux} are given in their circularly symmetric form for the
sake of simplicity, but GIM2D does consider the intrinsic (i.e. before
PSF convolution) ellipticity of the bulge by including bulge
ellipticity $e$ and position angle $\phi_b$ as additional parameters.

The GIM2D disk component is a simple exponential profile of the form:

\begin{equation}
	\Sigma(r) = \Sigma_{0} exp (-r/r_d).
	\label{disk} \end{equation}

\noindent $\Sigma_{0}$ is the face-on central surface brightness, and
$r_d$ is the semi-major axis exponential disk scale length.  The disk
is assumed to be infinitely thin.  The total flux in the disk is given
by:

\begin{equation}
	F_{\rm disk} = 2 \pi r_d^2 \Sigma_{0}.
	\label{diskflux} \end{equation}

\noindent The projected surface brightness distribution of the disk
inclined at any angle $i$ is calculated by integrating
Equation~\ref{disk} over the area in the face-on disk plane seen by
each projected pixel. The position angle $\phi_d$ of the disk is also
a fitting parameter.

Mosaicing dithered images with non-integer pixel shifts requires flux
interpolation between pixels and corrections for geometric distortion. 
For undersampled HST images such as ours, these processes severely 
distort the shape of the PSF and destroy the uncorrelated noise characteristics 
of theoriginal images. To circumvent these potential difficulties, GIM2D allows 
for fitting multiple, dithered individual images simultaneously. We use this mode 
for the NIC2 data. The WFPC2 images are also undersampled, but because 
our archival images have integer pixel shifts they are combined without information 
loss, and the combined image is analyzed with GIM2D.

HST point-spread-functions vary spatially across the image, and
additionally NICMOS PSFs vary temporally.  A PSF must therefore be
individually generated for each object of interest.  We use the
software package TINYTIM V5.0c \citep{krist93} to generate PSFs that
include both types of variation for each BCG position and observation
date (for NIC2).  Our NIC2 PSFs are 1\arcsecpoint9 $\times$
1\arcsecpoint9, and our WFPC2 PSFs are 2\arcsecpoint4 $\times$
2\arcsecpoint4. Both types of PSFs are oversampled by a factor of 5.
To compare directly to undersampled HST data, GIM2D creates a surface
brightness model on an oversampled grid, convolves it with the
corresponding oversampled PSF, shifts it to a subpixel position
specified by the values of the fitting parameter $dx$ and $dy$ and
rebins the resulting profile to the detector's resolution.

We use SExtractor segmentation images to isolate the BCGs from their
neighbors. As SExtractor performs source detection, it deblends
sources using flux multi-thresholding \citep{bertin96}.  The 
result of deblending are segmentation images in which
pixels belonging to the same object are all assigned the same flag
value, and sky background pixels are flagged by zeroes.  GIM2D fits
are performed on all pixels flagged as the target object {\it or}
background in the SExtractor segmentation images.  The rationale
behind not just fitting ``object'' pixels is that considering only the
pixels interior to the boundary between object and sky, which is
sharply delineated by the isophote corresponding to the detection
threshold, would exclude information contained in the
pixels below that threshold.

We fit the NIC2 BCG images with pure de Vaucouleurs ($n = 4$) bulge
and full de Vaucouleurs bulge + exponential disk models. Although the sky level can be a free
parameter in GIM2D, we fix the sky level to the value
determined from the curve of growth analysis of Section
\ref{analysis-cog} because of the relatively large angular extent of
the galaxies and the non-uniformity of the background.  The results
from the GIM2D NIC2 fits are shown in Table~\ref{nic2-sizes}. The
GIM2D half-light radii are measured along the semi-major axis (SMA) of
the BCG.  SMA radii can be converted to ``circular" radii according to
$r_{circ}$ = $r_{sma} \sqrt{1-e_{bcg}}$ where $e_{bcg}$ is the ellipticity
of the BCG. We list both SMA and circular radii in Table
\ref{nic2-sizes}.  Residual images are visually inspected to determine
which model (pure bulge or bulge+disk) yields a better fit.  The
better model is listed in Column 13 (``PB" = Pure Bulge, ``BD" =
Bulge+Disk and ``EQ" = Equally Good) of Table \ref{nic2-sizes}. Although
seven BCGs are fit better by a bulge+disk model, this result does {\it
not} imply that actual disks are present in those BCGs.  The residual
images suggest that an intrinsic change in ellipticity (i.e.  over
what one would expect from pure PSF smoothing) and position angle as a
function of radius may be responsible for the preference of a
bulge+disk model.

We fit the deeper and larger WFPC2 images with three different models:
a pure de Vaucouleurs bulge, a de Vaucouleurs bulge + exponential disk and a pure S\'ersic
of parameter $n$. We allow $n$ to vary between 1 and 6. Two sets of fits are
performed: one with the sky fixed to the value obtained from the
curve of growth analysis and one with the sky left as a free
parameter. The results are shown in Table~\ref{wfpc2-sizes}. There are
two lines per BCG in Table~\ref{wfpc2-sizes}.  The first and second
lines give the results of the fixed and floating background fits
respectively.  The half-light radius of the best model, as determined
through visual inspection of the residual images, is underlined for
each BCG. The advantage of using so many different models is the
ability to see the systematic effects on the measurement of the BCG
half-light radii introduced by the choice of a particular model.  The
significant disagreements between models arises because the half-light
radius depends on the measured total flux, which depends on the
details of the poorly constrained outer regions of the galaxy.
Table~\ref{wfpc2-sizes} lists six radii for each BCG, and the
variations among these six values reflect errors due to the
modeling of the sky and the flux in the outer wings of the galaxy.

\subsection{Elliptical Isophote Fitting}\label{analysis-isoellipse}

Elliptical isophote fitting provides yet another
way to measure BCG sizes and to demonstrate how bulge+disk model fits can
introduce a disk component to model a pure elliptical galaxy with
ellipticity and position angle radial variations. We use
the task ELLIPSE in the IRAF/STSDAS/ISOPHOTE package to fit isophotal
ellipses to the WFPC2 images of seven out of nine BCGs. We could not
acceptably fit MS1621 and MS0451. The ``BCG" in MS1621 is actually a
close pair of early-type looking galaxies, and the MS0451 BCG is
adjacent to another large galaxy.

In this fitting, we allow the isophote center,
ellipticity, and position angle to vary with radius. In contrast,
introducing radial variations is not practical in 
model fitting because of the large
number of degrees of freedom. For example, the GIM2D model has one center, one
ellipticity and one position angle per component (bulge or disk).  The
isophotal ellipse fitting 
results are presented in Table~\ref{wfpc2-ellipse-fits}.  The
Table presents the difference in ellipticity between the outer and
inner isophotes, $\Delta e$(ISO), the difference in ellipticity between
the outer (disk) component and the inner (bulge) component, $\Delta
e$(GIM2D), the difference in isophote position angle between the outer
and inner isophotes, $\Delta \phi $(ISO), the position angle difference
between the disk and bulge components, $\Delta \phi$(GIM2D), and the
pixel shifts in the isophote center between the outer and inner
isophotes, $\Delta x$ and $\Delta y$.  The close correlations between
$\Delta e$(ISO) and $\Delta e$(GIM2D) and between $\Delta \phi $(ISO)
and $\Delta \phi $(GIM2D) suggest that bulge+disk models fit some of
the BCG profiles better than bulge-only models because the extra
freedom in the two component models is being used to fit
the radial variations in the ellipticity
and position angle.

\subsection{NIC2 Curve-of-Growth Simulations}\label{analysis-sims}

To test how well we recover true half-light radii from NIC2 curves of
growth, we perform a set of simulations.  We choose five BCGs with
both NIC2 and WFPC2 images: MS1137, CL1322,
MS2053 and CL0016. We create three types of NIC2 simulations for each BCG.
For the first simulation, we take the NIC2 best-fit pure
$r^{1/4}$ GIM2D model image, and we insert it in another location on
the NIC2 image. This type of simulations is labeled as ``NIC2 $r^{1/4}$"
in Table~ \ref{nic2-sizes-sims}. For the second simulation (labeled as
``NIC2 B+D" in Table~ \ref{nic2-sizes-sims}), we take the NIC2 best-fit
bulge+disk GIM2D model, and we insert it in another location on the
NIC2. For the third simulation (labeled as ``WFPC2 $r^{1/4}$" in
Table~\ref{nic2-sizes-sims}), we use the half-light radius of the
best-fit WFPC2 model to create a pure $r^{1/4}$ model. We normalize
this model to have the measured flux, and we insert it in
the NIC2 image. We analyze all three types of 
simulations using the
same curve-of-growth procedure that we use to measure circular half-light
radii from the real data. See Table~\ref{nic2-sizes-sims} for the
results. No model fits are involved. We compare the radii
recovered from the curve-of-growth to the circular input values. $F_r$
is the ratio of the recovered and input radii, and $F_f$ is the ratio
of the recovered and input total fluxes. Using the results from all
three types of simulations, we find that the NIC2 half-light
radius is systematically underestimated by a mean of 20\%.
The cause of the problem is not simply the lack
of sky in the NIC2 images because
one of the galaxies with the largest WFPC2 half-light radii (MS1137)
is recovered well.

\section{Discussion}\label{discussion}

\subsection{Are High-Redshift BCGs in Dynamical Equilibrium?}

The presence of significant distortions indicating recent major mergers and/or 
an ellipticity distribution that is significantly different from the local BCG population 
would be clear morphological signs that high-redshift BCGs are not in dynamical 
equilibrium.  However, the surface brightness profiles of our 16 BCGs are quite regular 
(Figure 1). To generate the plotted profiles, we fit elliptical isophotes to the data using 
the IRAF task ELLIPSE. We use the best fitting model galaxy parameters listed in Table 
1 to compute the slope and zero-point of the corresponding PSF-deconvolved de 
Vaucouleurs profile. The semi-major axis radii are converted from arcsec to kpc assuming 
$H_{0} = 80$ km s$^{-1}$ Mpc$^{-1}$, $\Omega_{m} = 0.2$, and $\Omega_\Lambda = 0.0$. 
The BCG surface brightnesses have been corrected for cosmological $(1+z)^{4}$ dimming,
but are not K-corrected (K-corrections for an elliptical galaxy in the $H$-band are typically 
$\sim$0.2-0.4 mag for this redshift range). Figure 1 demonstrates that all but three of our 
BCGs are well described by a de Vaucouleurs profile beyond the resolution limit and out 
to at least $\sim$2 NIC2 $r_{e}$.  The notable exceptions are MS0302.5$+$1717, CL0016,  
and 1041+4626. The BCG MS0302.5$+$1717 is actually best-fit by a pure exponential disk 
profile. We note, however, that this BCG is selected directly from the NICMOS images and 
therefore its identification is less certain. The BCG CL0016 has a prominent dust lane, and 
the BCG 1041+4626 lies at the corner of a frame where the local background varies significantly.  
Additionally, we search for azimuthal asymmetry in the BCG profiles by subtracting the best 
fitting 2D galaxy model from the data. With the exception of the BCG in MS0302.5$+$1717, 
which is not well fit by an R$^{1/4}$ profile, and the BCG in CL0016, which shows a dust lane, 
we find that the BCG images do not have significant 2-dimensional residuals (the best fitting 
models typically yield $\chi^{2} \sim 1$). Consistent with this finding, our high redshift BCGs 
have similar ellipticities ($0.0 < \epsilon < 0.5$) to their local counterparts \citep{graham96}. 
We conclude that the BCGs exhibit no signs of being far from dynamical equilibrium.

\begin{figure*}
\plotone{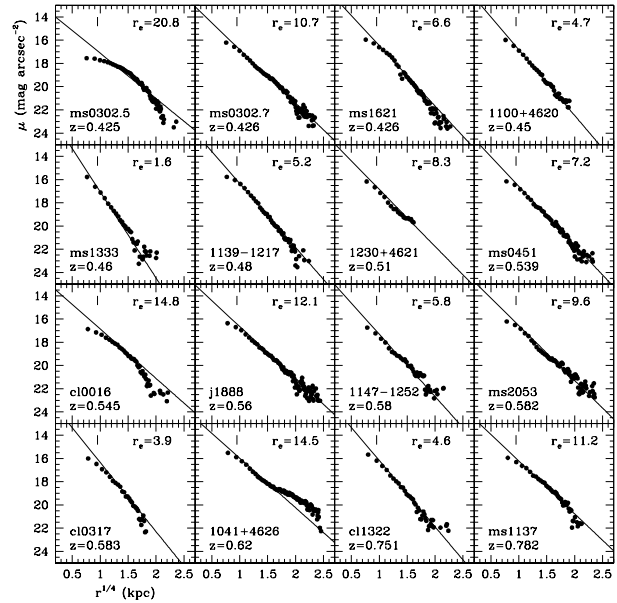}
\caption{The surface brightness profiles of our 16 BCGs ordered by
increasing redshift. The filled circles are the data and the line is the best fitting de Vaucouleurs
($r^{1/4}$) profile. The vertical mark in the upper left corner denotes
the angular resolution limit (i.e. FWHM of the point spread function) of the
NICMOS images. All but three of our 
BCGs are well described by a de Vaucouleurs profile beyond the resolution limit and out 
to at least $\sim$2 NIC2 $r_{e}$.  The notable exceptions are MS0302.5$+$1717, CL0016,  
and 1041+4626. The BCG MS0302.5$+$1717 is actually best-fit by a pure exponential disk 
profile, the BCG CL0016 has a prominent dust lane, and the BCG 1041+4626 lies at the corner 
of a frame where the local background varies significantly.\label{f1}}
\end{figure*}

\subsection{Comparison with Local BCG Sizes} The good agreement
between our BCG surface brightness profiles and a standard de
Vaucouleurs profile in conjunction with the lack of any significant
2-dimensional residuals argues against recent major accretion
events. However, cumulative minor merger events over time may manifest
themselves not by gross morphological distortions, but by the gradual
enlargement of the central galaxy. The simplest way to quantify the
sizes of these galaxies is through $r_{e}$. In Figure 2 we show the
distribution of semi-major axis $r_{e}$ from the GIM2D pure de Vaucouleurs bulge fits to our BCGs (\textit{black histogram}). To
avoid ``double-weighting" MS1621 because it is a binary BCG, 
we only include BCG MS1621a. 
The choice of BCG in MS1621 is inconsequential for this discussion
because both
galaxies of the ``BCG pair" have similar effective radii. We find
that our high redshift BCGs have a biweight-estimated effective radius
of $8.3\pm1.4$ kpc in the NICMOS images. 
We compare to the local sample of Graham et al.
(1996; \textit{shaded histogram}) who fit the surface brightness
profiles of BCGs from 119 Abell clusters at $z \le 0.05$ in the
$R$-band with a standard de Vaucouleurs profile (i.e. $n = 4$)
assuming $H_{0} = 80$ km s$^{-1}$ Mpc$^{-1}$.  Their low redshift BCGs
have a biweight-estimated effective radius (measured along the
semi-major axis as for our sample) of $16.1\pm1.7$ kpc which is $\sim
2$ times larger than that of our NIC2 high redshift sample. 
To quantitatively assess the significance of this difference in
$r_{e}$, we randomly select subsets of 16 BCGs from the \citet{graham96} 
local BCG sample and compute a biweight-estimated $r_{e}$ for
each subset. Using 500 Monte Carlo realizations of the high redshift
BCG sample drawn from the low redshift sample, we find that $< 3$\%
(13 of 500) of the realizations have a biweight-estimated $r_{e}$ less
than $8.3$ kpc. However, the result is qualitatively different if we
use the $F814W$ $r_e$'s of the 9 BCGs with WFPC2 images. The
biweight-estimated $r_e$ for that sample is $18.2 \pm 5.1$ kpc ---
larger than, and fully consistent with, the radii of the local sample.

\begin{figure*}
\plotone{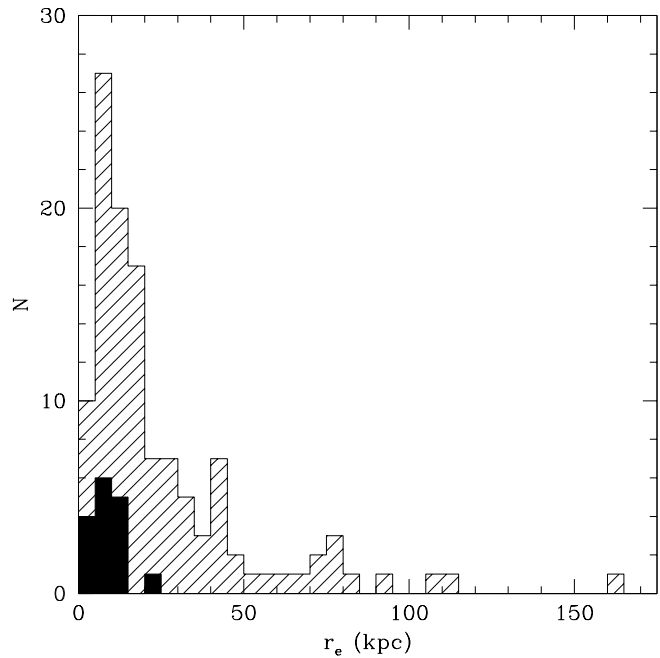}
\caption{The distribution of NIC2 semi-major axis effective radius, $r_{e}$, from the GIM2D pure de Vaucouleurs 
bulge fits to our high-redshift BCG sample (\textit{black histogram}) compared to the local Abell sample of Graham et al.  
(1996) (\textit{shaded histogram}).
\label{f2}}
\end{figure*}

\subsection{Are Optical and Infrared Radii Expected to be the Same?}
\begin{figure*}
\plotone{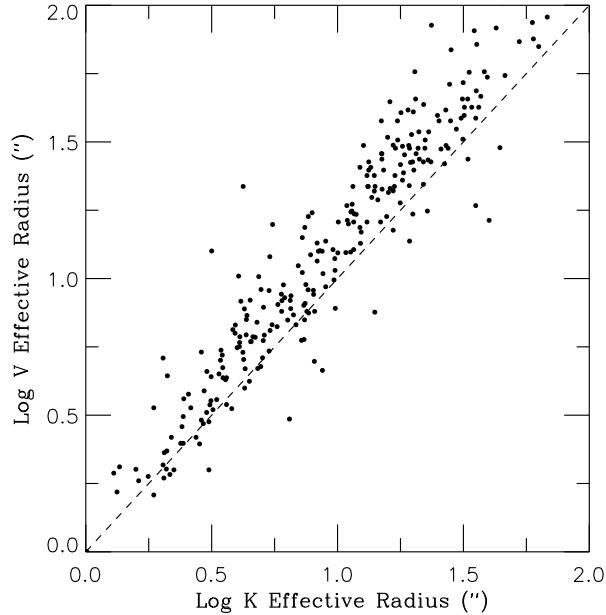}
\caption{Log $V$ effective radius and log $K$ effective radius in arcseconds for 273 E/S0 
galaxies from the sample of \citet{pahre99}. The dashed line is the one-to-one line. 
\label{f3}}
\end{figure*}

The discrepancy between the results for the NIC2 and WFPC2 radii of
our high-redshift BCG sample is either due to systematic errors in the
NIC2 measurements or to a real feature of BCGs. If systematic errors
are to blame, they would have to be substantially larger than the
simulations in Section \ref{analysis-sims} would suggest. Is it
possible that effective radii in the rest-frame range 4578$-$5676
\AA\thinspace\thinspace (F814W at $z$ = 0.6) are
systematically different from effective radii measured in the
rest-frame range 8750$-$11250 \AA\thinspace\thinspace (F160W at $z$ =
0.6)? Unfortunately, there are no large local samples of BCGs with
both optical and near-infrared photometric parameters.  However, a
tantalizing hint may come from the E/S0 sample of Pahre (1999). Pahre
measured global photometric parameters for a sample of 341 nearby
early-type galaxies in the near-infrared $K$-band. The majority (85\%)
of the galaxies in the Pahre sample reside in 13 rich clusters with
additional galaxies drawn from loose groups (12\%) and the field
(3\%).  The $K$-band sample was complemented with optical $V$-band
data from the literature. In Figure~\ref{f3} we plot the
$V$-band effective radius versus the $K$-band effective
radius for the 273 galaxies with both measurements. The dashed line is
the one-to-one line. It is immediately obvious that the $V$-band radii
are systematically larger than the $K$-band ones, and the discrepancy
increases with increasing radius. At log $r_e = 1.5$ in
$K$, the V radii are about 1.8 times larger than the K radii. As explained in 
Pahre et al. (1998), this difference is due to the presence of color gradients. Given
that BCGs are a distinct class of objects, it is unclear
whether BCG radii could exhibit the same type of wavelength dependence
as the E/S0's. However, if they do, then the discrepancy between our
NIC2 and WFPC2 radii would be explained, 
and the actual size evolution of the
BCGs between $z =0$ and 0.6 would be bounded by the WFPC2 images, which are
bluer in the rest-frame than the local sample, and the NICMOS images,
which are redder in the rest-frame than the local sample.

\subsection{Accretion Rates} Observations of BCGs show that the
relation between log $\sigma$ and $M$ (i.e., the Faber-Jackson
relation; 1976) is rather flat, i.e. BCGs are overluminous for their
velocity dispersions (Malumuth \& Kirshner 1981, 1985;
\citealp{oegerle91}). This result has been interpreted as support for
dissipationless merger scenarios for BCGs in which their masses,
luminosities, and effective radii increase, but their velocity
dispersions remain constant because $\sigma^{2} \propto M/R$. Because
our high redshift BCGs do not exhibit any signs of recent major
interactions, we presume that they are in dynamical equilibrium and that they 
follow the same scaling relations as local BCGs.  Consequently, if their mass 
scales directly with $r_{e}$, we would infer from the comparison of NIC2 and 
local effective radii that BCGs as a class have doubled in mass since $z \sim 
0.5-0.6$. Using the semi-analytic model predictions of both the Durham 
\citep{cole94,baugh96} and Munich \citep{kauffman93} groups, ABK98 present
the increase in BCG mass as a function of redshift. For CDM models
with $\Omega_{m} = 1$ our inferred mass accretion rate since $z \sim
0.5$ is in excellent agreement with the model predictions. For low
density CDM models with our adopted cosmology ($\Omega_m = 0.2$), they
predict slightly less mass accretion by BCGs since $z \sim 0.5$ (a
factor of $\sim$1.5 increase in mass), but given the large
observational uncertainties our findings are still in agreement.

On the other hand, the effective radii from the WFPC2 images show no
significant difference from the local BCG sample and would thus be
consistent with very little or {\it no} mass accretion since $z =
0.5-0.6$. If this is indeed the case, then we may be faced with the
continuing puzzle of a class of galaxies with sizes and magnitudes
that show little or no discernible evolution with redshift even though
their stellar population must at least be passively evolving (and
are observed to be passively evolving; \citealp{nelson01b}). Part of the
possible resolution to this puzzle lies
in the fact that evolution is tied to environment.  The majority of our
BCGs reside in clusters that fall in the high X-ray luminosity
category ($L_{x} > 2.3 \times 10^{44}$ ergs s$^{-1}$). As noted in the
introduction, the luminosities of high $L_{x}$ BCGs do match the
predictions of passive evolution models \citep{burke00,nelson01b}, which 
limits the possible amount of mass these BCGs accrete to less than a 
factor of 2. Despite their inherent systematic uncertainties, our 
NIC2/WFPC2 results are sufficient to rule out accretion rates higher 
than a factor of 2 in high $L_x$ BCGs, but our results cannot be used 
to place constraints on accretion rates below that level.

We emphasize that our inferred BCG mass accretion rates do not rest on
the assumption that the same galaxy identified as the BCG at high
redshift remains the BCG to the present day. Indeed the brightest
galaxy in any given cluster may very well change over time, but this
process is part of the evolution one aims to study. The value of BCGs
in evolutionary studies is their relatively unambiguous selection
criteria \textit{as a class} in both the data and simulations.

\section{Summary}\label{summary}

Using high angular resolution F160W NICMOS imaging, we measure the
luminosity profiles of 16 BCGs at $0.4 < z < 0.8$. The heterogeneous
sample is drawn from existing optical and x-ray surveys published in
the literature \citep{spinrad80,koo81,gunn86,couch91,gioia94,gonzalez01}. 
Archival WFPC2 F814W images were also available for 9 of the 16 BCGs. Our primary
conclusions are that:

\noindent 1) the NICMOS surface brightness profiles of high redshift
BCGs are well described by a de Vaucouleurs profile ($r^{1/4}$) out to
a radius of at least  $\sim 2$ NIC2 $r_{e}$,

\noindent 2) the NICMOS effective radii of our high redshift BCGs are,
on average, $\sim 2$ times smaller than those of local samples
\citep{graham96} for $H_{0} = 80$ km s$^{-1}$ Mpc$^{-1}$, $\Omega_{m} =
0.2$, and $\Omega_\Lambda = 0.0$. Using simple dynamical equilibrium arguments, 
this result suggests that BCGs
have increased in mass by a factor of $\sim 2$ since $z \sim 0.5$.

\noindent 3) the WFPC2 effective radii are fully consistent with those
of the local sample, which would suggest little or no mass
accretion. The discrepancy between the NICMOS and WFPC2 results
is either due to systematic errors in the
NICMOS measurements or to a real physical property of BCGs such as color gradients. If
systematic errors are to blame, they would have to be substantially
larger than the $\sim$ 20\% systematic radius error suggested by our simulations. Supporting our suggestion
that this difference is not due to a systematic error, we note
that the optical effective radii of local E/S0 galaxies are
larger than their near-infrared values due to color gradients. If the
effective radii of BCGs exhibit a similar wavelength dependence, it
would reconcile our NICMOS and WFPC2 radii. Given that our
WFPC2 and NICMOS data are both bluer and redder in the rest-frame
than the R-band local sample, the results from the two data sets
bound the possible evolution of $r_e$, and hence the accretion.

Although this sample marks a significant improvement in the quality of
data available for high redshift BCGs, our sample is too small to
fully resolve systematic uncertainties and to confidently constrain
mass accretion rates below a factor of 2. However, our work
demonstrates the feasibility of using high resolution imaging to
measure BCG structural parameters at $z>0.5$. With a sufficiently
large sample, complemented with some velocity dispersion measurements
to determine whether the scaling relations hold, such an approach has
the potential to definitively address the issue of mass accretion by
BCGs since $z \sim 1$.

\vskip 1in \noindent Acknowledgments: AEN would like to thank Kim-Vy
Tran (and her GIM2D Cookbook) for her help in understanding the
intricacies of GIM2D. AEN and AHG acknowledge funding from a NSF grant
(AST-9733111). AEN gratefully acknowledges financial support from the
University of California Graduate Research Mentorship Fellowship
program. AHG acknowledges funding from the ARCS Foundation and the CfA
postdoctoral fellowship program. DZ acknowledges financial support
from a NSF CAREER grant (AST-9733111), a David and Lucile Packard
Foundation Fellowship, and an Alfred P. Sloan Fellowship.  This work
was partially supported by NASA through grant number GO-07327.01-96A
from the Space Telescope Science Institute, which is operated by the
Association of Universities for Research in Astronomy, Inc., under
NASA contract NAS5-26555.


\clearpage
\begin{deluxetable}{lcccccc}
\tablewidth{0pt}
\tablecaption{HST NICMOS and WFPC2 BCG Image Datasets}
\tablehead{
\colhead{Cluster} & \colhead{$\alpha$} & \colhead{$\delta$} & \multicolumn{2}{c}{Total Integration} & 
\multicolumn{2}{c}{CADC Association} \\
\colhead{ID} & \colhead{(J2000.0)} & \colhead{(J2000.0)} & \multicolumn{2}{c}{Time (s)} & \multicolumn{2}{c}{Dataset ID}\\
\colhead{} & \colhead{} & \colhead{} & \colhead{F160W} & \colhead{F814W} & \colhead{F160W} & \colhead{F814W}\\
\colhead{(1)} & \colhead{(2)} & \colhead{(3)} & \colhead{(4)} & \colhead{(5)} & \colhead{(6)} & \colhead{(7)}
}
\startdata
CL0016+16 &  00:18:33.64 & +16:26:15.13 & 768 & 8400 & N42B38010 & U2C40101A\\
J1888.16CL &  00:56:56.78 & $-$27:40:31.26 & 768 & 8400 & N42B39010 & U2C40501A\\
MS0302.5+1717 & 03:05:18.19 & +17:28:23.35& 768 & \nodata & N42B48010 & \nodata \\
MS0302.7+1658 &  03:05:31.71 & +17:10:02.64 & 768 & 2400 & N42B49010 & U2UL0103A\\
CL0317+1521\tablenotemark{a} &  03:20:01.52 & +15:31:57.65 & 256 & 2500 & N42B40010 & U58J0101A\\
MS0451.6-0305 &  04:54:10.83 & $-$03:00:56.87 & 768 & 10400 & N42B50010 & U3060101A\\
1041$+$4626 &  10:41:03.81& +46:26:36.24 & 1056 & \nodata & N42B05010 & \nodata \\
1100$+$4620 &  11:00:57.40 & +46:20:37.88 & 768& \nodata & N42B15010 & \nodata \\
1139$-$1217 & 11:39:57.00 & $-$12:17:19.00 & 768 & \nodata & N42B12010& \nodata \\
MS1137.5+6625 & 11:40:23.86 & +66:08:18.47 & 1056 & 14400 & N42B52010 & U3060301A\\
1147$-$1252 & 11:47:17.30 & $-$12:52:09.00 & 768 & \nodata & N42B27010 & \nodata \\
1230$+$4621\tablenotemark{b} &  12:30:16.26 & +46:21:17.27 & 512 & \nodata & N42B07010 & \nodata \\
CL1322+3027 &  13:24:48.83 & +30:11:38.16 & 1056 & 14000 & N42B43010 & U2840501A\\
MS1333.3+1725 &  13:35:47.14 & +17:09:39.20 & 768 & \nodata & N42B53010 & \nodata \\
MS1621.5+2640 & 16:23:35.48 & +26:34:12.82 & 768 & 4600 & N42B55010 & U53B0701A\\
MS2053.7-0449 &  20:56:21.78 & $-$04:37:50.88 & 768 & 3200 & N42B56010 & U4F30601A 
\enddata

\label{hst-data}
\tablenotetext{a}{Two of the three dithered exposures for this object had pointing errors}
\tablenotetext{b}{One of the three dithered exposures for this object was lost due to data corruption.}
\end{deluxetable}

\begin{deluxetable}{lccrrrrrrrrrc}
\tablewidth{0pc}
\tablecaption{NIC2/F160W BCG Size Measurements }
\tablehead{
\colhead{BCG} & \colhead{$z$} & \multicolumn{2}{c}{Curve Of} & \multicolumn{4}{c}{Pure Bulge $r_e$} & \multicolumn{4}{c}{Bulge+Disk $r_{hl}$} & \colhead{Better}\\
\colhead{} & \colhead{} & \multicolumn{2}{c}{Growth $r_e$} & \multicolumn{2}{c}{sma} & \multicolumn{2}{c}{circ}  & \multicolumn{2}{c}{sma} & \multicolumn{2}{c}{circ} & \colhead{Model}\\
\colhead{} & \colhead{} & \colhead{\arcsec} & \colhead{kpc} & \colhead{\arcsec} & \colhead{kpc} &\colhead{\arcsec} & \colhead{kpc} & \colhead{\arcsec} & \colhead{kpc} &\colhead{\arcsec} & \colhead{kpc} & \colhead{}\\
\colhead{(1)} & \colhead{(2)} & \colhead{(3)} & \colhead{(4)} & \colhead{(5)} & \colhead{(6)} &\colhead{(7)} & \colhead{(8)} & \colhead{(9)} & \colhead{(10)} &\colhead{(11)} & \colhead{(12)} & \colhead{(13)}
}
\startdata 
CL0016  & 0.545 & 0.88 & 4.50 & 2.90 & 14.82 & 2.09 & 10.69 & 1.23 & 6.29 & 0.89 & 4.55 & BD\\
  &  & 1.21 & 6.19 & & & & & & & & &\\
  &  & 1.15 & 5.88 & & & & & & & & &\\
J1888 & 0.560 & 1.56 &  8.08 & 2.34 &  12.12 & 2.03 &  10.51 & 3.08 &  15.95 & 2.93 & 15.17 & EQ\\
 &  & 1.89 & 9.79 & & & & & & & & &\\
 &  & 1.53 & 7.92 & & & & & & & & &\\
MS0302.5 & 0.425 & 1.30 & 5.85 & 4.61 & 20.75 & 3.82 & 17.19 & 1.99 & 8.96 & 1.65 & 7.43 & BD\\
 &  & 1.62 & 7.29 &  & &  & &  & & & &\\
 &  & 1.51 & 6.80 &  & &  & &  & & & &\\
MS0302.7 & 0.426 & 1.51 & 6.80 & 2.38 & 10.72 & 2.09 & 9.42 & 1.61 & 7.26 & 1.41 & 6.35 & PB\\
 &  & 1.45 & 6.53 & & & & & && & &\\
 &  & 1.36 & 6.13 & & & & & && & &\\
CL0317 & 0.583 & 0.38 & 2.00 & 0.73 & 3.85 & 0.60 & 3.16 & 0.51 & 2.69 & 0.42 & 2.22 &BD \\
MS0451  & 0.539 & 0.93 & 4.73 & 1.41 & 7.17 & 1.06 & 5.39 & 1.10 & 5.59 & 0.82 & 4.17 &PB\\
  &  & 0.85 & 4.32 & & & & & & & & &\\
 &  & 0.73 & 3.71 & & & & & & & & &\\
1041$+$4626 & 0.620 & 2.96 & 16.04 & 2.67 & 14.47 & 2.40 & 13.01 & 2.68 & 14.53 & 2.36 & 12.79 &BD\\
 &  & 3.74 & 20.27 & & & & & & & & &\\
 &  & 3.91 & 21.19 & & & & & & & & &\\
1100$+$4620  & 0.450 & 1.00 & 4.64 & 1.02 & 4.73 & 1.00 & 4.64 & 1.23 & 5.71 & 1.21 & 5.62 &EQ \\
&  & 0.91 & 4.22 & & & & & & & & &\\
&  & 0.70 & 3.25 & & & & & & & & &\\
1139$-$1217 & 0.480 & 0.76 & 3.65 & 1.09 & 5.23 & 1.04 & 4.99 & 1.04 & 4.99 & 0.99 & 4.75 &BD\\
 &  & 0.67 & 3.22 & & & & & & & & &\\
 &  & 0.71 & 3.41 & & & & & & & & &\\
MS1137 & 0.782 & 1.77 & 10.51 & 1.89 & 11.21 & 1.84 & 10.92 & 1.91 & 11.34 & 1.86 & 11.04 &EQ\\
&  & 1.04 & 6.17 & & & & & & & & &\\
&  & 1.08 & 6.41 & & & & & & & & &\\
1147$-$1252 & 0.580 & 0.80 & 4.21 & 1.11 & 5.84 & 0.98 & 5.16 & 0.91 & 4.79 & 0.82 & 4.32 & EQ\\
 &  & 0.78 & 4.10 & & & & & & & & &\\
 &  & 0.93 & 4.89 & & & & & & & & &\\
1230$+$4621\tablenotemark{a} & 0.510 & 2.00 & 9.90 & 1.68 & 8.32 & 1.58 & 7.82 & 1.96 & 9.70 & 1.72 & 8.52 &BD\\
 &  & 1.62 & 8.02 & & & & & & & & &\\
CL1322 & 0.751 & 0.61 & 3.57 & 0.78 & 4.56 & 0.57 & 3.34 & 0.76 & 4.45 & 0.55 & 3.22 &EQ \\
 &  & 0.95 & 5.56 & & & & & & & & &\\
 &  & 0.54 & 3.16 & & & & & & & & &\\
MS1333 & 0.460 & 0.50 & 2.35 & 0.33 & 1.55 & 0.33 & 1.55 & 0.47 & 2.21 & 0.47 & 2.21 & BD\\
 &  & 0.58 & 2.72 & & & & & & & & & \\
 &  & 0.54 & 2.54 & & & & & & & & & \\
MS1621a  & 0.426 & 0.88 & 3.97 & 1.46 & 6.58 & 1.28 & 5.77 & 1.05 & 4.73 & 0.99 & 4.46 &PB\\
 &  & 0.76 & 3.42 & & & & & && & &\\
 &  & 0.80 & 3.60 & & & & & && & &\\
MS1621b &  0.426 & 0.81 & 3.65 & 1.36 & 6.13 & 1.33 & 5.99 & 1.09 & 4.91 & 0.90 & 4.06 &BD\\
 &  & 0.75 & 3.38 & & & & & && & &\\
 &  & 0.88 & 3.96 & & & & & && & &\\
MS2053 & 0.582 & 1.46 & 7.70 & 1.82 & 9.59 & 1.75 & 9.22 & 1.77 & 9.33 & 1.71& 9.01 & PB\\
 &  & 1.43 & 7.54 & & & & & & & & & \\
 &  & 1.17 & 6.17 & & & & & & & & &
\enddata
\tablenotetext{a}{One of the three dithered exposures for this object was lost due to data corruption.}
\label{nic2-sizes}
\end{deluxetable}

\begin{deluxetable}{lrrrrrrrrrrrrrrc}
\tablewidth{0pc}
\tablecaption{WFPC2/F814W BCG Size Measurements }
\tablehead{
\colhead{BCG} & \multicolumn{2}{c}{Curve Of} & \multicolumn{4}{c}{Pure Bulge $r_e$} & \multicolumn{4}{c}{Bulge+Disk $r_{hl}$} & \multicolumn{4}{c}{Pure S\'ersic $r_{e}$} & \colhead{S\'ersic}\\
\colhead{} & \multicolumn{2}{c}{Growth $r_e$} & \multicolumn{2}{c}{sma} & \multicolumn{2}{c}{circ}  & \multicolumn{2}{c}{sma} & \multicolumn{2}{c}{circ} & \multicolumn{2}{c}{sma} & \multicolumn{2}{c}{circ} 
& \colhead{Index $n$}\\
\colhead{} & \colhead{\arcsec} & \colhead{kpc} & \colhead{\arcsec} & \colhead{kpc} &\colhead{\arcsec} & \colhead{kpc} & \colhead{\arcsec} & \colhead{kpc} &\colhead{\arcsec} & \colhead{kpc} & \colhead{\arcsec} & \colhead{kpc} &\colhead{\arcsec} & \colhead{kpc} & \colhead{}\\
\colhead{(1)} & \colhead{(2)} & \colhead{(3)} & \colhead{(4)} & \colhead{(5)} &\colhead{(6)} & \colhead{(7)} & \colhead{(8)} & \colhead{(9)} &\colhead{(10)} & \colhead{(11)} & \colhead{(12)} & \colhead{(13)} &\colhead{(14)} & \colhead{(15)} & \colhead{(16)}
}
\startdata 
CL0016  	& 3.20  & 16.36 & \underline{8.50} & 43.46 & 6.96 & 35.59 & 5.67 & 28.99 & 4.96 & 25.36 & 6.04 & 30.88 & 4.98 & 25.46 & 3.3\\
              	&           & & 8.98 & 45.91 & 7.35 & 37.58 & 5.02 & 25.67 & 4.39 & 22.44 & 6.46 & 33.02 & 5.29 & 27.05 & 3.4\\
J1888 	& 2.76  & 14.29 & 4.43 & 22.94 & 4.01 & 20.77 & \underline{3.27} & 16.93 & 3.14 & 16.26 & 6.66 & 34.49 & 6.03 & 31.23 & 4.9\\
           	&           & & 4.26 & 22.06 & 3.86 & 19.99 & 2.48 & 12.84 & 2.38 & 12.32 & 6.60 & 34.18 & 5.98 & 30.97 & 4.9\\
MS0302.7 & 2.76  & 12.44 & 4.77 & 21.49 & 4.22 & 19.01 & 3.30 & 14.87 & \underline{2.69} & 12.12 & 5.34 & 24.06 & 4.72 & 21.27 & 4.3\\
                   &           & & 5.08 & 22.89 & 4.50 & 20.28 & 3.28 & 14.78 & 2.68 & 12.08 & 7.02 & 31.63 & 6.24 & 28.12 & 4.7\\
CL0317 	&  0.53 & 2.80 & 0.91 & 4.80 & 0.66 & 3.48 & 0.60 & 3.16 & 0.51 & 2.69 & 0.60 & 3.16 & 0.44 & 2.32 & 2.5\\
           	&           & & 0.93 & 4.91 & 0.67 & 3.53 & \underline{0.58} & 3.06 & 0.50 & 2.64 & 0.59 & 3.11 & 0.43 & 2.27 & 2.4\\
MS0451  	& 2.87  & 14.60 & 4.19 & 21.31 & 3.54 & 18.00 & \underline{3.75} & 19.07 & 2.33 & 11.85 & 6.79 & 34.53 & 5.76 & 29.29 & 5.1\\
                 	&           & & 3.85 & 19.58 & 3.27 & 16.63 & 3.68 & 18.72 & 2.29 & 11.65 & 6.14 & 31.23 & 5.21 & 26.50 & 4.9\\
MS1137 	& 2.83  & 16.80 & 4.49 & 26.52 & 4.31 & 25.58 & 3.53 & 20.95 & 3.02 & 17.93 & 5.51 & 32.71 & 5.28 & 31.34 & 4.5\\	 	&           & & 4.56 & 27.07 & 4.56 & 27.07 & \underline{3.57} & 21.19 & 3.06 & 18.16 & 4.95 & 29.38 & 4.75 & 28.20 & 4.3\\
CL1322 	&  1.17 & 6.85 & \underline{1.15} & 6.73 & 1.08 & 6.32 & 2.36 & 13.81 & 2.31 & 13.52 & 1.95 & 11.41 & 1.83 & 10.71 & 5.4\\
	 	&           & & 1.08 & 6.32 & 1.02 & 5.97 & 1.10 & 6.44 & 1.08 & 6.32 & 1.95 & 11.41 & 1.83 & 10.71 & 5.4\\
MS1621a  & \nodata & & 2.66 & 11.99 & 2.51 & 11.31 & 3.65 & 16.45 & 3.15 & 14.19 & 3.25 & 14.64 & 3.05 & 13.74 & 4.8\\
                   &                & & 1.96 & 8.83 & 1.84 & 8.29 & \underline{2.01} & 9.06 & 1.73 & 7.80 & 1.44 & 6.49 & 1.34 & 6.04 & 3.2\\
MS1621b  & \nodata & & 2.07 & 9.33 & 2.03 & 9.15 & \underline{2.24} & 10.09 & 1.58 & 7.12 & 1.68 & 7.57 & 1.65 & 7.43 & 3.4\\
                   &                & & 1.96 & 8.83 & 1.93 & 8.70 & 1.74 & 7.84 & 1.23 & 5.54 & 1.05 & 4.73 & 1.02 & 4.60 & 2.5\\
MS2053 	& 1.77  & 9.33 & \underline{2.32}  & 12.23 & 2.05 & 10.81 & 2.33 & 12.28 & 2.02 & 10.65 & 2.19 & 11.54 & 1.92 & 10.12 & 3.8\\
		&           & & 2.28  & 12.02 & 2.00 & 10.54 & 1.52 & 8.01 & 1.32 & 6.96 & 1.93 & 10.17 & 1.69 & 8.91 & 3.6
\enddata
\label{wfpc2-sizes}
\end{deluxetable}

\begin{deluxetable}{lcccccccc}
\tablewidth{0pc}
\tablecaption{WFPC2 BCG Isophotal Ellipse Fitting}
\tablehead{
\colhead{BCG} & \colhead{Fitting Range} & \colhead{$r_e$} & \colhead{$\Delta e$} & \colhead{$\Delta e$} & \colhead{$\Delta \phi$} & \colhead{$\Delta \phi$} & \colhead{$\Delta x$} & \colhead{$\Delta y$}\\
\colhead{} & \colhead{in $r^{1/4}$} & \colhead{sma} & \colhead{(ISO)} & \colhead{(GIM2D)} & \colhead{(ISO)} & \colhead{(GIM2D)} & \colhead{pixels} & \colhead{pixels}\\
\colhead{(1)} & \colhead{(2)} & \colhead{(3)} & \colhead{(4)} & \colhead{(5)} & \colhead{(6)} & \colhead{(7)} & \colhead{(8)} & \colhead{(9)}
}
\startdata 
CL0016		& 1.1 $-$ 2.8 & 8.35 & +0.14 & +0.19 & $-$23 & $-$31 & +3 & $-$2\\
J1888		& 1.0 $-$ 2.1 & 5.71 & +0.10 & \thinspace\thinspace\thinspace\thinspace\thinspace 0.00 & $-$25 & $-$30 & \thinspace\thinspace\thinspace\thinspace\thinspace0 & \thinspace\thinspace\thinspace\thinspace\thinspace 0\\
MS0302.7 	& 1.1 $-$ 3.0 & 7.02 & +0.15 & +0.39 & +20 & +10 & \thinspace\thinspace\thinspace\thinspace\thinspace 0 & $-$3\\
CL0317		& 1.5 $-$ 2.3 & 0.28 & +0.10 & +0.25 & \thinspace\thinspace\thinspace  \thinspace\thinspace\thinspace\thinspace\thinspace 0 & +\thinspace\thinspace\thinspace 2 & \thinspace\thinspace\thinspace\thinspace\thinspace0 & +2\\
MS1137 		& 1.1 $-$ 2.6 & 6.32 & +0.14 & +0.26 & $-$30 & $-$47 & \thinspace\thinspace\thinspace\thinspace\thinspace0 & \thinspace\thinspace\thinspace\thinspace\thinspace0\\
CL1322		& 1.1 $-$ 2.3 & 1.31 & $-$0.05 & \thinspace\thinspace\thinspace\thinspace\thinspace 0.00 & $-$20 & +20 & +2 & +2\\
MS2053		& 1.3 $-$ 2.6 & 2.35 & $-$0.10 & $-$0.20 & +24 & +60 & \thinspace\thinspace\thinspace\thinspace\thinspace 0 &\thinspace\thinspace\thinspace\thinspace\thinspace 0\\
\enddata
\label{wfpc2-ellipse-fits}
\end{deluxetable}

\begin{deluxetable}{llccccc}
\tablewidth{0pc}
\tablecaption{NIC2 BCG Size Measurement Simulations}
\tablehead{
\colhead{BCG} & \colhead{Model} & \multicolumn{2}{c}{Input $r_{hl}$} & \colhead{Recovered} & \colhead{$F_r$} & \colhead{$F_f$}\\
\colhead{} & \colhead{Type} & \colhead{sma} & \colhead{circ} & \colhead{COG $r_{hl}$} & \colhead{} & \colhead{}\\
\colhead{(1)} & \colhead{(2)} & \colhead{(3)} & \colhead{(4)} & \colhead{(5)} & \colhead{(6)} & \colhead{(7)}
}
\startdata 
CL0016	& NIC2 $r^{1/4}$ & 2.89 & 2.26 & 0.95 & 0.42 & 0.51\\
		& NIC2 B+D & 1.28 & 0.92 & 1.02 & 1.10 & 0.89\\
		& WFPC2 $r^{1/4}$ & 4.86 & 3.80 & 1.12 & 0.30 & 0.38\\
MS0302.7 & NIC2 $r^{1/4}$ & 2.38 & 2.26 & 1.73 & 0.77 & 0.82\\
		& NIC2 B+D & 1.61 & 1.42 & 1.72 & 1.21 & 1.07\\
		& WFPC2 $r^{1/4}$ & 3.46 & 3.28 & 3.97 & 1.21 & 1.16\\
MS1137 & NIC2 $r^{1/4}$ & 1.89 & 1.81 & 1.53 & 0.85 & 0.88\\
	          & NIC2 B+D & 1.92 & 1.86 & 1.56 & 0.84 & 0.89\\
		 & WPFC2 $r^{1/4}$ & 3.72 & 3.55 & 2.90 & 0.82 & 0.95\\
CL1322	& NIC2 $r^{1/4}$ & 0.78 & 0.57 & 0.46 & 0.81 & 0.66\\
		& NIC2 B+D & 0.76 & 0.55 & 0.42 & 0.77 & 0.63\\
		& WPFC2 $r^{1/4}$ & 2.70 & 1.97 & 0.61 & 0.31 & 0.28\\
MS2053	& NIC2 $r^{1/4}$ & 1.82 & 1.62 & 1.42 & 0.88 & 0.98\\
		& NIC2 B+D & 1.77 & 1.70 & 1.61 & 0.94 & 1.04\\
		& WFPC2 $r^{1/4}$ & 2.52 & 2.24 & 1.68 & 0.75 & 0.90\\
\enddata
\label{nic2-sizes-sims}
\end{deluxetable}

\end{document}